\documentclass[runningheads]{svmult}

\usepackage{makeidx}   % allows index generation
\usepackage{graphicx}  % standard LaTeX graphics tool
\usepackage{subeqnar}  % subnumbers individual equations
\usepackage{multicol}  % used for the two-column index
\usepackage{physprbb}  % modified textarea for proceedings,
\makeindex             % used for the subject index
\begin{document}
\title*{Cosmic acceleration from effective forces?}
\toctitle{Cosmic acceleration from effective forces?}
\titlerunning{Cosmic acceleration from effective forces?}
\author{Dominik J.~Schwarz\inst{1}
\and Winfried Zimdahl \inst{2}
\and Alexander B.~Balakin\inst{3}
\and Diego Pav\'{o}n\inst{4}}
\authorrunning{Dominik J.~Schwarz et al.}
\institute{Institut f\"ur Theoretische Physik, Technische Universit\"at Wien,\\ 
     Wiedner Hauptstra\ss e 8-10, A-1040 Wien, Austria
\and Fachbereich Physik, Universit\"at Konstanz, \\
     Postfach M678, D-78457 Konstanz, Germany
\and Department of General Relativity and Gravitation, Kazan State University,\\
     RU-420008 Kazan, Russia
\and Departamento de F\'{\i}sica, Universidad Aut\'{o}noma de Barcelona,\\ 
     E-08193 Bellaterra (Barcelona), Spain}
\maketitle              % typesets the title of the contribution

\begin{abstract}
Accelerated expansion of the Universe may result from an anti-frictional force
that is self-consistently exerted on cold dark matter (CDM). Cosmic 
anti-friction is shown to give rise to an effective negative pressure of the 
cosmic medium. While other models introduce a component of dark energy 
besides ``standard'' CDM, we resort to a phenomenological one-component model
of CDM with internal self-interactions. We demonstrate how the dynamics of
the $\Lambda$CDM model may be recovered as a special case of cosmic
anti-friction \cite{ZSBP}. 
\end{abstract}

Measurements of supernovae of type Ia at high redshifts provide evidence 
that the expansion of the Universe accelerates \cite{R98,P99,R00} and 
there are first observational indications that this acceleration sets in at a 
redshift below $z \sim  1$ \cite{R01}. Observational 
aspects and some cosmological consequences of high-z supernovae have been 
discussed by R.~Kirshner at this conference \cite{K}. In our contribution we 
point out that other explanations besides introducing a cosmological constant
or another scalar (``quintessence'') field are possible.

Recently, the spatial geometry of the Universe has been measured by means 
of the temperature fluctuations in the cosmic microwave background (CMB)
\cite{Boomerang,CMB}. Based on a seven parameter fit the  
Boomerang team finds for the spatial curvature $\Omega_k = 0.03\pm0.06$ 
(value for weak priors) \cite{Boomerang}, which is consistent with a spatially 
flat Universe. 

The combination of CMB and high-z supernova data shows that dust (vanishing 
pressure $P$) alone cannot make up all of the mass in the Universe. An obvious 
solution to this problem is provided by a cosmological constant, 
parameterised by $\Omega_\Lambda$. However, this solution is plagued by heavy 
theoretical problems: Why is $\Omega_\Lambda$ so small? Why can we see it 
just know? While the first question is probably one of fundamental physics, 
an answer to the second question may come from cosmology. To avoid this 
``coincidence problem'', models with a new scalar field (``quintessence'' 
\cite{CDS}) have been introduced. These models can be characterised by the 
density of the scalar field, $\Omega_\phi$, and its equation of state 
$w_\phi \equiv P_\phi/\rho_\phi$. 

In the following we present an alternative explanation of CMB and high-z
supernova data in which one does not introduce another dark component besides
CDM (see \cite{ZSBP} for more details). Instead we add an extra force that 
acts on CDM. For non-relativistic CDM particles the most general force which 
is compatible with the cosmological principle was shown to reduce to
\begin{equation}
\vec{F} = - B(z) m \vec{v},
\end{equation}   
where $m$ is the mass of the particles and $\vec{v}$ their peculiar velocity.
This force is directed parallel or anti-parallel to the motion of the
individual CDM particles. The quantity $B(z)$, which has the dimension of
an inverse time, obviously plays the role of a coefficient of (anti-)friction. 
Macroscopically, the action of a force of this type generates an effective 
viscous pressure. Thus it is not allowed to describe CDM by a perfect fluid in
that case. The exact expression for the dynamic pressure in the 
presence of (anti-)frictional forces reads 
\begin{equation}
P = \frac B H \rho,
\end{equation} 
which gives rise to an accelerated expansion of the Universe if $B < -H/3$, 
thus in the case of anti-friction. In our calculation we have assumed that 
the CDM particle distribution is invariant with respect to elastic collisions
(the collision term in the kinetic equation vanishes), which implies that 
a negative pressure is related to the phenomenon of particle production. 

\begin{figure}[b]
\includegraphics[width=.47\textwidth]{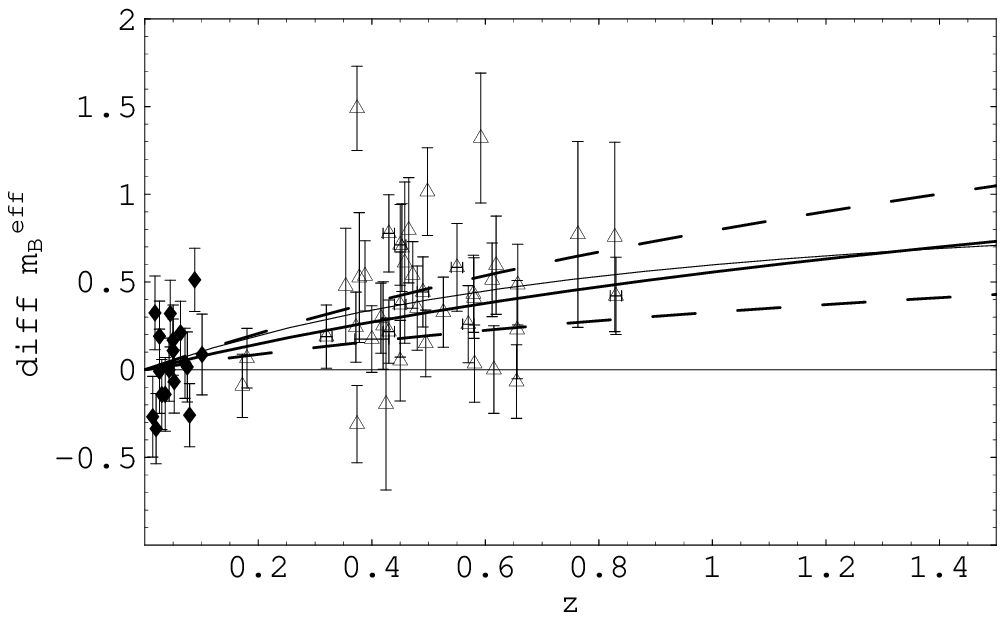}
\hfill \includegraphics[width=.47\textwidth]{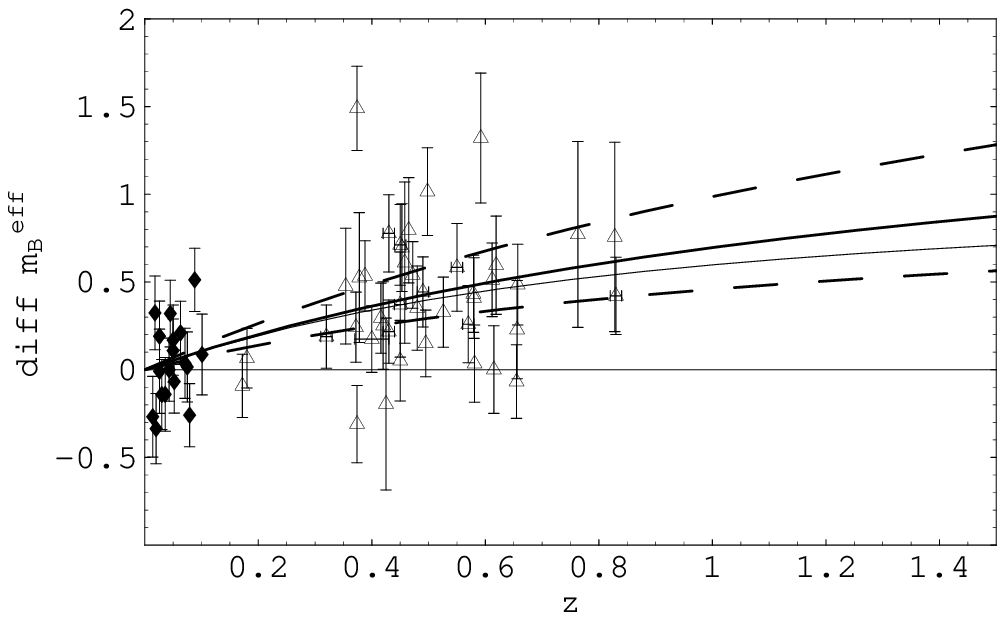}
\caption[]{Differences of the magnitudes with respect to a 
$(\Omega_{\rm M},\Omega_\Lambda) = (1,0)$ universe versus redshift. The data 
points are taken from \cite{P99}. The thin line denotes the $(\Omega_{\rm M},
\Omega_\Lambda) = (0.3,0.7)$ universe. For the lhs figure we assume 
$B = - \nu H$ and plot the predictions for $\nu = 0.7,0.5,0.3$ (thick lines 
from top to bottom). For the figure on the rhs we assume $B = - \nu H_0$ with 
$\nu = 0.9,0.7,0.5$ (thick lines from top to bottom).}
\label{fig1}
\end{figure}

\begin{figure}[t]
\centerline{\includegraphics[width=.47\textwidth]{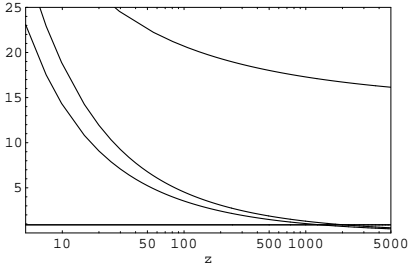}}
\caption[]{The angular scale under which the Hubble horizon at redshift $z$
would be observed today. The horizontal line is the angular scale, $0.9^\circ$,
of the first acoustic peak in the CMB temperature spectrum. The curves 
show our three models $-B \propto H, H_0, 1/H$ (from top to bottom), 
respectively. The first model is excluded by the CMB data, the last one is
dynamically equivalent to $\Lambda$CDM.}
\label{fig2}
\end{figure}

Even without having a microscopic model for cosmic anti-friction at hand we 
can check this idea by a comparison with the data. We need to make an ansatz 
for $B(z)$. It is natural to assume $B = - {\cal O}(H_0)$, since $H_0$ is the 
only rate that is distinguished by cosmology. In \cite{ZSBP} we investigated 
three different models, which all fit the supernova data.
Figure \ref{fig1} shows the prediction for two different models. 
With the help of the CMB data we can rule out one of our models. The two 
other models seem to be consistent, as is shown in figure \ref{fig2}.

It turns out that our third ansatz $B = - \nu H_0^2/H$, where $\nu$ is a 
constant, is dynamically equivalent to the $\Lambda$CDM model, 
since $P = - \nu \rho_0$ in that case. This ansatz fits both the CMB and 
the supernova data and predicts an onset of acceleration at $z_{\rm acc} 
\approx 0.7$, but with the important difference that there is no separate 
dark energy component.
This degeneration with $\Lambda$CDM may be resolved by investigating 
large scale structure or clusters. Naively, one might think that 
cluster data \cite{CB} immediately rule out $\Omega_{\rm M} = 1$. However,
one should keep in mind that $\Omega_{\rm cluster}$ typically is obtained 
under the assumption that CDM is a perfect fluid. In general relativity
the gravitating mass is given by $\rho + 3 P$, thus if there were a similar 
negative dynamic pressure at cluster scales, the matter densities from clusters
could be underestimated. 
%%%%%%%%%%%%%%%%%%%%%%%%%%%%%%%%%%%%%%%%%%%%%%%%%%%%%%%%%%%%%%%%%%%%%%%%% 

We have shown that effective anti-frictional forces could explain
the accelerated expansion of the Universe. An interesting hint 
for finding a microscopic model of cosmic anti-friction might be 
that it is the cosmological scale $1/H_0$ that gives rise to the correct order 
of magnitude, which seems to suggest a gravitational mechanism. 

D.J.S. thanks the Austrian Academy of Sciences for financial support. This 
work was also supported by the Deutsche Forschungsgemeinschaft and NATO.

\end{document}